\documentclass[10pt,aps,prl,twocolumn,amsmath,amssymb,nofootinbib,superscriptaddress]{revtex4-2}

\usepackage{mathtools}		
\usepackage{xcolor}
\usepackage[caption=false]{subfig}
\usepackage{titlesec}
\usepackage{float}

\usepackage[colorlinks=true,
            citecolor=red,
            linkcolor=blue,
            urlcolor=violet,
            filecolor=cyan,
            backref=false]{hyperref}

\makeatletter
\renewcommand\section[1]{%
  \par
  \addvspace{1.0ex}%
  \noindent\hspace*{\parindent}%
  \textit{#1}---%
  \ignorespaces
}
\makeatother

\newcommand{\lie}{\pounds}
\newcommand\bs{\boldsymbol}
\renewcommand{\cosh}{\operatorname{ch}}

\renewcommand{\tanh}{\operatorname{th}}

\newcommand\feq{\mathrel{\phantom{=}}}

\newcommand{\sgn}{\operatorname{sgn}}

\begin{document}
\title{Rotating near-horizon extreme geometries in quadratic gravity}

\author{Breno L. Giacchini}
\email{breno.giacchini@matfyz.cuni.cz}
\affiliation{
{\small Institute of Theoretical Physics, Faculty of Mathematics and Physics, Charles University, V Hole{\v s}ovi{\v c}k{\'a}ch 2, 180 00 Prague 8, Czech Republic}
}

\author{Ivan Kol\'a\v{r}}
\email{ivan.kolar@matfyz.cuni.cz}
\affiliation{
{\small Institute of Theoretical Physics, Faculty of Mathematics and Physics, Charles University, V Hole{\v s}ovi{\v c}k{\'a}ch 2, 180 00 Prague 8, Czech Republic}
}

\date{\today}

\begin{abstract}
Rotating near-horizon extreme solutions of quadratic gravity are analyzed combining power series expansions and numerical calculations. Restricting to geometries with symmetries of the near-horizon extreme Kerr black hole (i.e., with the $\mathrm{AdS_2}$-structure), spherical horizon topology, and equatorial reflection symmetry, we introduce conformal coordinates 
simplifying the field equations of Einstein--Weyl gravity, i.e., quadratic gravity with vanishing scalar curvature. 
Employing the Frobenius analysis, we classify all power series solutions expanded around the equator as well as the poles, and obtain the recurrence relations. With the help of numerical analysis, we study fine-tuning of the free parameter to satisfy the global constraints on regular near-horizon extreme geometries. Computations of the horizon area, horizon scalar curvature, and rotational scalar reveal strong horizon deformations in some Bachian branches. Moreover, the absence of regular Bachian near-horizon geometries above a finite horizon area suggests an upper bound on the size of the corresponding extremal rotating black holes.
\end{abstract}

\maketitle

\section{Introduction}
While static black-hole solutions in quadratic gravity have been extensively studied \cite{Lu:2015psa,Lu:2015cqa,Podolsky:2018pfe,Podolsky:2019gro}, exact rotating solutions remain elusive. Such solutions are important for understanding how higher-curvature corrections modify the Kerr solution. A natural route toward understanding properties of spinning black holes is provided by near-horizon extreme geometries. The field equations become tractable when restricted to the commonly occurring structure of two-dimensional anti-de Sitter ($\mathrm{AdS}_2$) spacetime, which appears, for example, in the vacuum solution of general relativity (GR) known as the near-horizon extreme Kerr (NHEK) geometry \cite{Bardeen:1999px}. The rotating near-horizon extreme geometries have been thoroughly studied in GR (see \cite{Kunduri:2013gce} for a review), and were also considered Einstein--Born--Infeld gravity \cite{Hale:2025urg}, Einsteinian cubic gravity \cite{Cano:2019ozf}, and scalar-tensor theories with topological curvature couplings \cite{Lam:2026grz}. 
In this Letter, we construct, for the first time, families of exact vacuum solutions of this type in quadratic gravity.

\section{Quadratic gravity}
Quadratic gravity represents the simplest and most extensively studied higher-derivative extension of GR. It is described by the Lagrangian
\begin{equation}
    L= \gamma R+\beta R^2-\alpha C_{abcd}C^{abcd}\;,
\end{equation}
where $R$ and $C_{abcd}$ are the Ricci scalar and Weyl tensor, respectively, and $\gamma$, $\beta$, and $\alpha$ are coupling constants. Curvature-squared terms make quadratic gravity perturbatively renormalizable, at the cost of additional degrees of freedom \cite{Stelle:1976gc}. Besides the massless graviton, the theory propagates a massive spin-2 mode and a scalar mode around flat spacetime. The former is tachyonic for ${\omega=\alpha/\gamma<0}$ (as in the asymptotic safety scenario~\cite{Benedetti:2009gn,Hamada:2017rvn}); hence, we refer to ${\omega<0}$ (${\omega>0}$) as \textit{tachyonic (non-tachyonic)} models. 
The ${R=0}$ sector of quadratic gravity coincides with Einstein--Weyl gravity (${\beta=0}$), whose vacuum field equations are
\begin{equation}\label{eq:FE}
\begin{gathered}
    R_{ab}=4\omega B_{ab}\;,
    \qquad
    B_{ab}=\left(\nabla^c\nabla^d+\tfrac12 R^{cd}\right)C_{acbd}\;,
\end{gathered}
\end{equation}
where $R_{ab}$ is the Ricci tensor, and $B_{ab}$ is the Bach tensor, which is traceless
, symmetric
, conserved
, and conformally covariant
.

\section{Near-horizon metric ansatz}
Although the metric form of the most general near-horizon limit of any extremal rotating black hole is well known \cite{Kunduri:2007vf,Kunduri:2008tk,Kunduri:2008rs,Kunduri:2013gce}, we focus on a simpler subclass of metrics that admit the same symmetries as the NHEK geometry. 
As shown in 
Appendix~\ref{app:metric}, the general metric ansatz admitting the NHEK symmetries can be written as
\begin{gather}
    \bs{g} = c(r)\bs{q} + a(r)\bs{\alpha}^2 + \tfrac{\bs{\mathrm{d}}r^2}{a(r)}\;, \label{eq:metric_ac}
    \\
    \bs{q}=- \rho^2 \bs{\mathrm{d}}\tau^2 +\tfrac{\bs{\mathrm{d}}\rho^2}{\rho^2-1}\;, \quad \bs{\alpha}=\bs{\mathrm{d}}\chi {-}2n\sqrt{\rho^2{-}1}\bs{\mathrm{d}}\tau \;, \label{eq:qalph}
\end{gather}
where $\bs{q}$ is the metric on $\mathrm{AdS}_2$ with unit radius, and ${n}$ is an arbitrary non-zero parameter. Clearly, this general metric class describes an $S^1$ fibration over a 3-dimensional quotient, whose induced metric contains an $\mathrm{AdS}_2$ sector with $\mathrm{so}(2,1)$ symmetry. Notice that $r$ covers the space of orbits, while ${(\tau,\rho,\chi)}$ are coordinates on individual orbits. Furthermore, we assume ${a>0}$ and ${c>0}$ to arrive at a stationary Lorentzian metric. 
The NHEK solution is regained by setting ${a={(n^2-r^2)}/{(r^2+n^2)}}$ and ${c=r^2+n^2}$. To later interpret the metrics \eqref{eq:metric_ac} as near-horizon geometry of extremal rotating black holes, we change the coordinates on the orbits ${(\tau,\rho,\chi)\to(u,w,\phi)}$ (see Appendix~\ref{app:metric}) so that
\begin{equation}\label{eq:met_wu}
    \bs{q}=-w^2\bs{\mathrm{d}}u^2+\bs{\mathrm{d}}u\vee\bs{\mathrm{d}}w\;, \quad \bs{\alpha}=\bs{\mathrm{d}}\phi-2nw\bs{\mathrm{d}}u\;,
\end{equation}
where $\bs{\partial}_{\phi}$ generates the ${\mathrm{U(1)}}$ symmetry of the $S^1$ fibers. The class of spacetimes \eqref{eq:metric_ac} [or \eqref{eq:met_OmegaH} below] is a Kundt metric 
of Petrov type D and Segre type \{(1,1),11\} \cite{Kunduri:2013gce,Colleaux:2026} (see Appendix~\ref{app:metric} for further details and generalizations to other symmetries). The horizon is located at ${w=0}$ \cite{Kunduri:2013gce}. The induced metric on its arbitrary section $\bs{\gamma}$ together with the rotation 1-form $\bs{h}$ comprise all horizon data; in coordinates ${(\phi,r)}$ they are given by
\begin{equation}\label{eq:horizondata}
    \bs{\gamma} =a(r)\bs{\mathrm{d}}\phi^2 + \tfrac{\bs{\mathrm{d}}r^2}{a(r)}\;, 
    \quad\bs{h} =-\tfrac{2na(r)}{c(r)}\bs{\mathrm{d}}\phi-\!\tfrac{c'(r)}{c(r)}\bs{\mathrm{d}}r\;.
\end{equation}

Since we are interested in near-horizon geometries that may arise as limits of rotating extremal black holes, we impose further global assumptions. In particular, we demand the horizon metric $\bs{\gamma}$ to have spherical topology. As explained in \cite{Kunduri:2008rs}, this translates into ${a>0}$ in ${r\in(r_-,r_+)}$ with the roots only at the endpoints ${a(r_{\pm})=0}$ and finite ${c(r_\pm)>0}$. Here, $r$ is a latitudinal coordinate 
interpolating between north and south poles of the symmetry axis. Assuming differentiability and non-degenerate axis then implies finite ${a'(r_-)>0}$, ${a'(r_+)<0}$. Expanding ${a=a'(r_\pm)(r-r_\pm) +\mathcal{O}\big((r-r_\pm)^2\big)}$ and introducing ${\varrho=2\sqrt{(r-r_\pm)/a'(r_\pm)}}$, the horizon metric takes the form ${\bs{\gamma}=\bs{\mathrm{d}}\varrho^2+\big(a'(r_\pm)/2\big)^2 \varrho^2\bs{\mathrm{d}}\phi^2+\mathcal{O}(\varrho^4)}$. Denoting the $\phi$-periodicity by ${2\pi\nu>0}$, the absence of conical deficits requires ${a'(r_\pm)=\mp {2}/{\nu}}$. Thus, any ${a'(r_{-})=-a'(r_{+})>0}$ can be accommodated by a suitable choice of $\nu$. The symmetry axis is then automatically regular also outside the horizon \cite{Colleaux:2026}. Moreover, a sufficient condition is that $a$ is an even function. Then ${r_{\pm}=\pm r_{*}}$, ${r_{*}>0}$, and we only need to satisfy
\begin{equation}\label{eq:cond1}
    a=-\tfrac{2}{\nu} (r-r_{*}) +\mathcal{O}\big((r-r_*)^2\big)\;,
\end{equation}
for some ${\phi}$-periodicity parameter ${\nu>0}$. In what follows, we further restrict to full equatorial reflection symmetry,
\begin{equation}\label{eq:cond2}
    a(r)=a(-r)\;, \quad c(r)=c(-r)\;,
\end{equation}
where we used the residual freedom of \eqref{eq:metric_ac} under a constant shift of $r$ to place the equator at ${r=0}$.

\section{Conformal form and field equations} Due to the presence of the Bach tensor in the field equations of Einstein--Weyl gravity \eqref{eq:FE}, it is useful to further rewrite the metric in a conformal form where one function appears only as the conformal factor and the other only enters the seed metric. This can be achieved by a change of the orbit-space coordinate ${r\to\bar{r}}$, so that ${\bs{\mathrm{d}}r=\Omega^2\bs{\mathrm{d}}\bar{r}}$, together with the redefinitions ${a(r)=H(\bar{r}(r))\Omega^2(\bar{r}(r))}$ and ${c(r)=\Omega^2(\bar{r}(r))}$, where ${H>0}$ and ${\Omega>0}$ are functions of $\bar{r}$; the resulting metric is
\begin{equation}\label{eq:met_OmegaH}
    \bs{g}=\Omega^2(\bar{r})\left[\bs{q}+H(\bar{r}) \bs{\alpha}^2+\tfrac{\bs{\mathrm{d}}\bar{r}^2}{H(\bar{r})}\right]\;.
\end{equation}
At the level of the field equations, this leads to the reduction of the number of derivatives; specifically, \eqref{eq:EW_FE} below, will contain only up to second derivatives of $\Omega$.

The conditions \eqref{eq:cond1} and \eqref{eq:cond2} can be rewritten in terms of $\Omega$ and $H$. Evenness of $c$ implies evenness of the derivative of the transformation ${\bar{r}'(-r)=\bar{r}'(r)}$, which upon fixing the integration constant ${\bar{r}(0)=0}$ means ${\bar{r}(-r)=-\bar{r}(r)}$. Together with the evenness of $a$, it results in the evenness of $\Omega$ and $H$. Hence, ${\bar{r}\in(-\bar{r}_*,\bar{r}_*)}$, where ${\bar{r}_*>0}$ parametrizes the poles corresponding to the roots ${H(\pm\bar{r}_*)=0}$. Taking the derivative of $a$ expressed using $\Omega$ and $H$, ${a'=(H\Omega^2)'/\Omega^2=H'+2H\Omega'/\Omega}$, we arrive at the equivalent conditions,
\begin{equation}
\begin{gathered}
    H=-\tfrac{2}{\nu} (\bar{r}-\bar{r}_{*}) +\mathcal{O}\big((\bar{r}-\bar{r}_*)^2\big)\;,
    \\
    \Omega(\bar{r})=\Omega(-\bar{r})\;, \quad H(\bar{r})=H(-\bar{r})\;,
\end{gathered}
\end{equation}
where ${0<\Omega(\bar{r}_*)<\infty}$ and ${-\infty<H'(\bar{r}_*)<0}$.

In the conformal coordinates \eqref{eq:met_OmegaH} [either with ${(\tau,\rho,\chi,\bar{r})}$ or ${(u,w,\phi,\bar{r})}$] the field equations \eqref{eq:FE} reduce to the set
\begin{subequations}\label{eq:EW_FE}
    \begin{eqnarray}
        \Omega \Omega'' - 2 \Omega'^2 - n^2 \Omega^2 & = & - \tfrac{\omega }{3 H}\mathcal{B}_1\;,
        \label{eq:EW_FE-tr}
        \\
        \Omega \Omega' H' + 3 \Omega'^2 H + \Omega^2 ( 1 - n^2 H) & = & -\tfrac{\omega}{3} \mathcal{B}_2\;,
        \label{eq:EW_FE-rr}
    \end{eqnarray}
\end{subequations}
where $\mathcal{B}_1$ and $\mathcal{B}_2$ are two independent components of the Bach tensor,
\begin{equation}\label{eq:Bach12}
\begin{aligned}
        \mathcal{B}_1 & =  H \big[H^{(4)}+4 n^2 \big(5 H''-4\big)+64 n^4 H\big]\;,
        \\
        \mathcal{B}_2  &=  H^{(3)} H' -\!\tfrac{1}{2} H''^2+\!2 n^2 \big(5 H'^2+\!16 n^2 H-\!8\big) +\!2\;.\!\!\!\!
\end{aligned}
\end{equation}
Although this follows directly from the above, the trace of the field equations (${R=0}$) reads
\begin{equation}\label{eq:EW_trFE}
    6 \Omega'' H +6 \Omega' H'+\Omega \left(H''-2 n^2 H+2\right) = 0\;.
\end{equation}

\section{NHEK branch}
The single-function metric subclass can be characterized covariantly by specializing to Segre type \{(1,1),(11)\}~\cite{Colleaux:2026}, or equivalently by
\begin{equation}\label{eq:covSF}
     S_{ab} S^{ab} -  \tfrac{3S^{ab} S_{cd}\left(C_{aefb} C^{cefd} - \tilde{C}_{aefb} \tilde{C}^{cefd}\right)}{C_{abcd} C^{abcd}}=0\;, 
\end{equation}
where ${S}_{ab}$ denotes the trace-free part of the Ricci tensor and  ${\tilde{C}_{abcd} = \frac12 \varepsilon_{abef}C^{ef}{}_{cd}}$ is the dual Weyl tensor. The condition \eqref{eq:covSF} reduces to ${\Omega \Omega''-2 \Omega'^2-n^2 \Omega^2=0}$, which is equivalent to ${\mathcal{B}_1=0}$ upon using the field equation \eqref{eq:EW_FE-tr}. It is solved by ${\Omega=\Omega_0\sec(n\bar{r})}$ for some constant ${\Omega_0>0}$, where we used the shift freedom in $\bar{r}$ to render the function even and positive around ${\bar{r}=0}$. Assuming this subclass, Eq.~\eqref{eq:EW_trFE} alone already restricts the even solutions to ${H={\big(c_1 \cos (2 n \bar{r})+c_1-1\big)\cos ^2(n \bar{r})}/{n^2}}$; imposing the full field equations  \eqref{eq:EW_FE} then yields ${c_1=1}$, i.e., 
the NHEK metric, for which ${\mathcal{B}_1=\mathcal{B}_2=0}$.
Notice an emergent scaling freedom 
\begin{equation}\label{eq:scalingfreedom}
    {n\to Sn}\;, 
    \quad
    {\bar{r}\to \bar{r}/S}\;, 
    \quad
    {\chi\to S\chi}\;, 
    \quad
    \text{(or ${\phi\to S\phi}$)}\;,
\end{equation}
with ${S}$ being a non-zero constant, which allows us to achieve ${\bar{r}_*=1}$ [i.e., ${\bar{r}\in(-1,1)}$], by setting ${n=\pi/4}$.

\section{Frobenius analysis}
More general solutions can be obtained using the Frobenius method, assuming that the metric functions are expandable as power series around a certain point $\bar{r}=\bar{r}_0$,
\begin{equation}\label{eq:OmF-series}
\Omega = \sqrt{\vert \omega\vert} \bar{\Delta}^N \mathop{\textstyle\sum}\limits_{i=0}^\infty  n^{N+i} f_i \bar{\Delta}^i \;, 
\;\; {H} = \bar{\Delta}^P \mathop{\textstyle\sum}\limits_{i=0}^\infty n^{P-2+i} h_i \bar{\Delta}^i \;, 
\end{equation}
where $\bar{\Delta} = \bar{r} - \bar{r}_0$ and $f_0,h_0\neq 0$ to ensure that the leading
exponents are $N$ and $P$ (yet to be determined). For convenience, we factored $\sqrt{\vert \omega\vert}$ in the definition of $\Omega$, so that the equations~\eqref{eq:EW_FE} for the coefficients $f_i$ and $h_i$ depend only on the sign of $\omega$.
(Although~\eqref{eq:OmF-series} introduces an artificial dependence on $\omega$ in the NHEK branch, in this degenerate case it is removable by the substitution $f_0 = \Omega_0 / \sqrt{\lvert\omega\rvert}$.) As will become clear below, the dependence on the parameter ${n \neq 0}$ can also be resolved by defining the series coefficients 
as above. A solution like~\eqref{eq:OmF-series} is said to belong to the class $[N,P]$. The analysis of the indicial equations reveals that the only possible classes of solutions are $[0,0]$, $[0,1]$, $[1,0]$, and $[-1,2]$ \cite{GiacchiniKolarInPrep}. According to the physical interpretation of the metric, solutions of type $[0,0]$ correspond to expansions around a generic point (including the equator, ${\bar{r}=0}$), whereas those $[0,1]$ contain expansions around the poles (${\bar{r}=\pm\bar{r}_*}$). 
The latter two classes give  expansions around a zero and a singularity of $\Omega$ 
and, therefore, do not concern the geometries we study here.

Recurrence relations for the coefficients $f_i$ and $h_i$ of the two types of expansions can be obtained using the Frobenius method [see Eqs.~\eqref{eq:recurr00f}--\eqref{eq:h201app} in Appendix~\ref{app:recur} 
for the explicit formulas]. Remarkably, thanks to the coordinates used and parameterization of the metric functions, both sets of relations are independent of $n$. In fact, these solutions also have the scaling freedom~\eqref{eq:scalingfreedom}, and $n$ can be tuned so that the roots of ${H}$ always satisfy ${\bar{r}_*=1}$. Also, the only dependence on $\omega$ is through $\sgn\omega$.

\begin{figure}[t]
    \centering
    \includegraphics[width=1\linewidth]{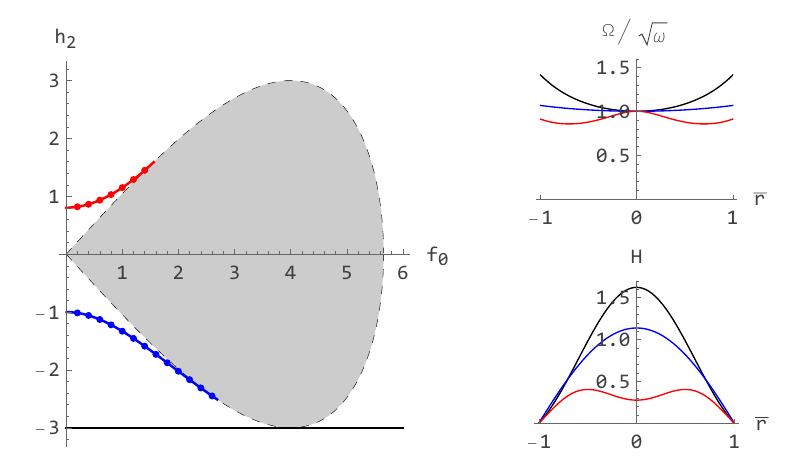}
    \caption{\textit{Left:} Branches of solutions in the parameter space for the non-tachyonic model. The (nontrivial) Bachian branches of $h_{0-}$ and $h_{0+}$ are depicted in blue and red, respectively, with dots indicating fine-tuned values and solid lines their interpolations. The (trivial) NHEK branch corresponds to the black line at ${h_2=-3}$; the forbidden region where $h_{0\pm}$ is not real is colored in gray. \textit{Right:} Solution for $f_0=1$ in each one of the three disconnected branches; more solutions are shown in Fig.~\ref{fig:Fig2-3}.}
    \label{fig:Fig1}
\end{figure}

Although a generic solution in $[0,0]$ can have at most five free physical parameters among $\{ f_0, f_1, h_0, h_1,  h_2, \allowbreak  h_3 \}$, the requirement of evenness yields $f_{1}=h_1=h_3=0$ for expansions around the equator ${\bar{r}_0=0}$. Therefore, $f_0$ and $h_2$ can be taken as free parameters, and $h_0$ defines two potential branches of solutions [see \eqref{eq:const00}],
\begin{equation}\label{h0pm}
    h_{0\pm} =  \tfrac{16+3\sgn\omega f_0^2  \pm \sgn\omega \sqrt{9 f_0^4 -288 \sgn\omega f_0^2 +256 h_2^2 }}{64}\;,
\end{equation}
provided ${9 f_0^4 -288 \sgn\omega f_0^2 +256 h_2^2 \geq 0}$. For ${\sgn\omega=+1}$ (non-tachyonic theory), this condition determines a bounded region on the parameter space where no solution exists (Fig.~\ref{fig:Fig1}). No similar restriction occurs for ${\sgn\omega=-1}$. 
Note that the NHEK solution with free parameter $f_0>0$ follows from the choice $h_0=1$, ${h_2=-3}$; it is contained in the branch $h_{0+}$ of the tachyonic theory, while in the non-tachyonic theory it splits between  branches ${h_{0+}}$ for ${0<f_0<4}$ and ${h_{0-}}$ for ${f_0\geq 4}$.

To study the global solutions with nontrivial Bach tensor, we combine expansions around the equator ${\bar{r}_0=0}$, generic points ${\bar{r}_0\in(-\bar{r}_*,0)\cup(0,\bar{r}_*)}$ and the poles ${\bar{r}_0=\pm\bar{r}_*}$, and also numerical routines in Wolfram Mathematica. 
A number of scenarios incompatible with the properties defining the rotating near-horizon extreme geometry can occur for generic values of the parameters ${f_0=\Omega(0)/\sqrt{|\omega|}}$ and ${h_2=H''(0)/2}$. For example, $\Omega$ may diverge or have a root, and $H$ may not have roots. Instead of presenting a numerical study of the whole parameter space, in what follows, we describe three families of global solutions that we identified for which, for a given $f_0$ it seems necessary to fine tune $h_2$.

\section{Bachian branches for non-tachyonic theory}
\begin{figure}
    {
    \includegraphics[width=0.95\linewidth]{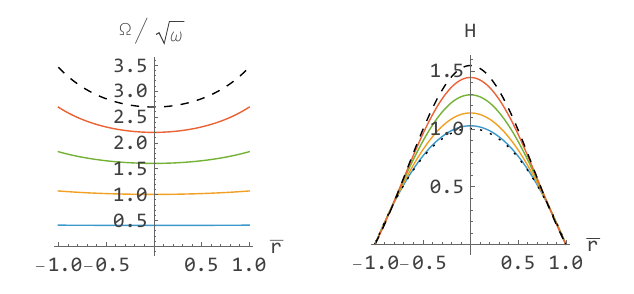}}
    \\
    {
    \includegraphics[width=0.95\linewidth]{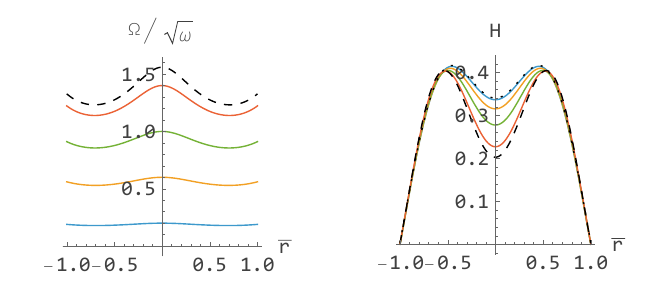}}
    \caption{Metric functions for the Bachian solutions of the non-tachyonic theory in the branches $h_{0-}$ (top) and $h_{0+}$ (bottom) for several values of ${f_0=\Omega(0)/\sqrt{|\omega|}}$. The dashed curves indicate the limiting solution for ${f_0\approx 2.69}$ (top) or ${f_0\approx 1.56}$ (bottom). $H$ approaches the dotted curve for $f_0\ll1$.}
    \label{fig:Fig2-3}
\end{figure}

We identified two branches of solutions for this case, one for each sign option in~\eqref{h0pm}. The branch related to ${H(0)=h_{0-}/n^2}$ requires ${H''(0)=2h_2<0}$ and exists for ${0<f_0 \lesssim 2.69}$, 
while the branch related to $h_{0+}$ has ${h_2>0}$ and ${0<f_0 \lesssim 1.56}$,
see Fig.~\ref{fig:Fig1}.
For larger values of $f_0$, the curves ${(f_0,h_2(f_0))}$ in the parameter space would intersect the region where~\eqref{h0pm} is not real. In principle, the curves could continue after this forbidden region, but we did not identify any such continuation. The solutions, for different values of $f_0$, are depicted in Fig.~\ref{fig:Fig2-3}. (Unless otherwise stated, all graphs show numerical solutions validated by the method described below.)

Our strategy to identify the solutions was the following: First, for a given $f_0$ we  numerically integrate the field equations~\eqref{eq:EW_FE} from the equator to the poles for different values of $h_2$. 
For smaller (larger) values of $\vert h_2\vert$, the numerical solutions exhibit a blow-up of $\Omega$ towards decreasing (increasing) values as ${\bar{r}\to\bar{r}_*}$.
The fine tuning that results in the curve ${h_2=h_2(f_0)}$ consists in identifying the point where this sign flip takes place.

\begin{figure}
    \centering
    \includegraphics[width=0.9\linewidth]{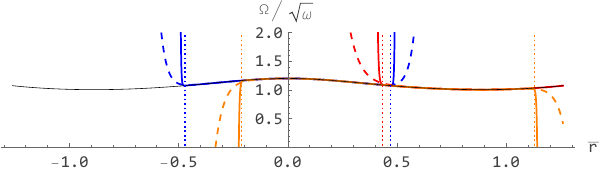}
    \\
    \vspace{0.4cm}
    \includegraphics[width=0.9\linewidth]{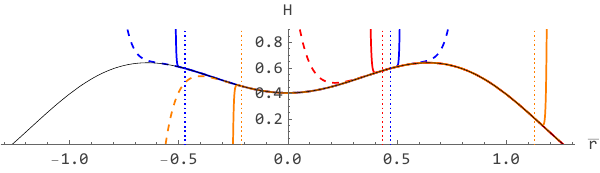}
    \caption{Series expansions for the solution in the branch $h_{0+}$ with $n=1$ and $f_0=1.2$ around $\bar{r}=0$ (blue), $\bar{r}=0.46$ (orange), and $\bar{r}=\bar{r}_*$ (red). Dashed lines depict truncation at 20 terms and solid lines at 200 terms;
    dotted vertical lines indicate the intervals of convergence. The numerical solution is represented by the thin black curve.}     
    \label{fig:Fig4}
\end{figure}

Despite fine-tuning, the poles (${\bar{r}=\pm\bar{r}_*}$) are singular points of the differential equations~\eqref{eq:EW_FE}, where numerical solutions tend to be ill-behaved. To ensure that they can be trusted all the way to the poles, we compare them with the analytical expansions. Specifically, using the fine-tuned $h_2$ for expansions around the equator, we estimate $\bar{r}_*$  and the values of $\Omega(\bar{r}_*)$, $H'(\bar{r}_*)$, and $H''(\bar{r}_*)$. These quantities are enough to reconstruct the series expansion of type $[0,1]$ around the poles [see Eqs.~\eqref{eq:recurr01f:f1}--\eqref{eq:h201app}]. 
Since the convergence radius of the expansions ${[0,0]}$ is typically smaller than $\bar{r}_*$ (especially for the ${h_2>0}$ branch), 
it might be necessary to combine the expansions around several intermediate points ${\bar{r}_0\in(0,\bar{r}_*)}$ until the behavior of the solution at ${\bar{r}=\bar{r}_*}$ can be estimated with confidence; for an example, see Fig.~\ref{fig:Fig4}.

\section{Bachian branch for tachyonic theory}
\begin{figure}[t]
    \centering
    \includegraphics[width=1\linewidth]{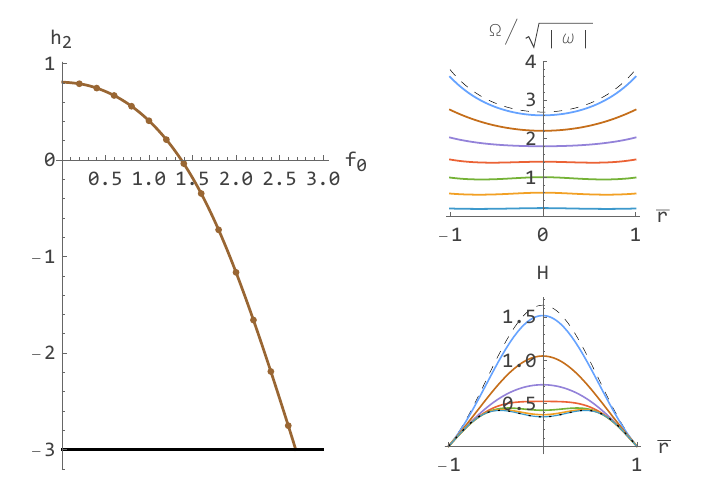}
    \caption{\textit{Left:} Branches of solutions of the tachyonic model in the parameter space. The Bachian branch for $h_{0-}$ (brown) meets the NHEK one (black) for ${f_0\approx 2.69}$. \textit{Right:} Solutions for several values of ${f_0=\Omega(0)/\sqrt{|\omega|}}$. The dashed curves indicate the limiting solution for ${f_0\approx 2.69}$, which is equivalent to NHEK. $H$ approaches the dotted curve for small values of~$f_0$.}
    \label{fig:Fig5}
\end{figure}

Applying the same procedure described previously, we identified a single Bachian branch of solutions, related to the choice $h_{0-}$ in~\eqref{h0pm}. Like the analogous branch in the non-tachyonic model, it exists for ${0<f_0 \lesssim 2.69}$ (Fig.~\ref{fig:Fig5}). At ${f_0 \approx 2.69}$ it connects to the NHEK branch, and we found no continuation beyond. This family of solutions can have negative and positive values of $h_2$, with a range ${-3 < h_2 \lesssim 0.81 }$. Solutions for several values of $f_0$ are shown in Fig.~\ref{fig:Fig5}.

\section{Horizon properties of the solutions}
To characterize the solution branches, we examine three quantities associated with the horizon: the horizon area $\mathcal{A}_{\mathcal H}$, the scalar curvature $R_{\mathcal{H}}$ of the induced metric on the horizon section, and the rotational scalar $\Upsilon$. 
They follow directly from the horizon data~\eqref{eq:horizondata} and are defined in Appendix~\ref{app:geometric}. Regarding the horizon area, all Bachian solutions are typically associated with an area smaller than that of NHEK, except for very small values of the parameter $f_0$, see Fig.~\ref{fig:BachArea}. 
Interestingly enough, the existence of a maximum value for the parameter $f_0$ translates into an upper bound for the horizon area. Together with the experimental bounds on the coupling $\alpha$ of the Weyl-squared term~\cite{Giacchini:2016nta} obtained from torsion balance experiments~\cite{Kapner:2006si}, we estimate this upper bound to be of order $10^{-5}\,\text{m}^2$ for the $h_{0-}$ branches, and $10^{-6}\,\text{m}^2$ for the $h_{0+}$ one, i.e., such extreme Bachian black holes would be microscopic.

For the non-tachyonic model, the Bachian $h_{0+}$ branch has a more distorted horizon compared to NHEK, as the scalar curvature $R_{\mathcal{H}}$ of the induced metric is such that ${R_{\mathcal{H}}|_{\bar{r}=0}}<0$, 
while NHEK and the  $h_{0-}$ branch have ${R_{\mathcal{H}}|_{\bar{r}=0}}>0$, see Fig.~\ref{fig:AP}. Unlike NHEK, both branches have ${R_{\mathcal{H}}|_{\pm \bar{r}_*}>0}$ for small values of $f_0$, but it becomes negative as $f_0$ increases. In the tachyonic theory, the horizon can vary from highly distorted (for small $f_0$) to a small perturbation of the NHEK one, Fig.~\ref{fig:AP}. The latter may be captured by perturbative solutions \cite{SvarcInPrep}.

The rotational scalar $\Upsilon$ follows a similar pattern, being more distorted for the $h_{0+}$ branch, see Fig.~\ref{fig:AP}. Indeed, while $\Upsilon$ for NHEK has two critical points and changes sign once, for the $h_{0+}$ branch it has four critical points and three changes of sign. In the branch $h_{0-}$ of the non-tachyonic theory, $\Upsilon$ changes sign only once, but may have no critical point if $f_0$ is small enough. $\Upsilon$ for the branch $h_{0-}$ of the tachyonic theory can have one or three changes of sign and two, three or four critical points.

\section{Closing remarks} Our analysis provides the first systematic study of rotating near-horizon extreme geometries in quadratic gravity, revealing properties of extremal rotating configurations that are otherwise difficult to access in higher-derivative gravity. 
Apart from recovering the NHEK branch as the only solution within the single-function subclass, we showed evidence that the spherical horizon topology together with equatorial symmetry and regularity at the poles seem to require a delicate fine-tuning of expansion data. 

In the non-tachyonic theory (Figs.~\ref{fig:Fig1}--\ref{fig:Fig2-3}), we identified two Bachian solution branches, both disconnected from the NHEK branch in the parameter space. 
Compared to the static spherically symmetric solutions \cite{Lu:2015cqa,Lu:2015psa,Podolsky:2018pfe,Podolsky:2019gro,Bonanno:2019rsq,Silveravalle:2022wij}, which admit at most one Bachian branch, rotation appears to split it into up to two separate branches with small values of the parameter $f_0$ and limited horizon area. This suggests that eventual rotating extremal black holes of which these solutions are the limits are microscopic, similar to the static ones~\cite{Silveravalle:2022wij}.
In the tachyonic theory (Fig.~\ref{fig:Fig5}), we identified a single Bachian branch, which is connected to the NHEK branch in the parameter space. This contrasts to the various branches of static spherically symmetric black holes found in~\cite{Bonanno:2019rsq}.
Remark that our solutions may be locally related, via a double Wick rotation~\cite{Colleaux:2025uiw}, to the hyperbolic version of the solutions of~\cite{Chen:2024hsh}. Determining the conserved quantities and analyzing the corresponding near-horizon thermodynamics \cite{Hajian:2013lna} could further clarify whether regular solutions exist for all physically reasonable conserved quantities and, consequently, whether rotating solutions with Cauchy horizons exist \cite{Hale:2025urg}.

\begin{acknowledgments}
We thank Robert \v{S}varc for valuable discussions and careful reading of the manuscript, and Vojt\v{e}ch Pravda and Alena Pravdov\'a for insights concerning the convergence radius of the infinite series.
We acknowledge support from the Primus grant PRIMUS/23/SCI/005 of Charles University. B.L.G. also acknowledges support from the Charles University Research Center grant UNCE24/SCI/016.
\end{acknowledgments}

\clearpage



\setcounter{equation}{0}
\setcounter{figure}{0}
\setcounter{table}{0}

\renewcommand{\theequation}{E\arabic{equation}}
\renewcommand{\thefigure}{E\arabic{figure}}
\renewcommand{\thetable}{E\arabic{table}}


\setcounter{section}{0}
\renewcommand{\thesection}{\Alph{section}}

\let\endmattersection\section
\renewcommand{\section}[1]{%
  \refstepcounter{section}%
  \par
  \addvspace{1.0ex}%
  \noindent\hspace*{\parindent}%
  \textit{Appendix~\thesection: #1}---%
  \ignorespaces
}


\section{Metric ansatz and transformations} \label{app:metric}%
The Killing vectors $\bs{X}_i$ (${i=1,\dots,4}$) of NHEK correspond to [4,3,9] in the notation of \cite{Hicks:thesis}. It is fully characterized by a specific Lorentzian Lie algebra-subalgebra pair whose abstract algebra corresponds to $\mathrm{so(2,1)\oplus \mathbb{R}}$ and the isotropy subalgebra is F14 (in the classification of \cite{Patera:1974zd}). Following \cite{Hicks:thesis,Frausto:2024egp,Colleaux:2025uiw}, the general spacetime invariant under this infinitesimal group action, ${\lie_{\bs{X}_i}\bs{g}=0}$, can be described by the metric in coordinates ${(\tau,\rho,\chi,r)}$,
\begin{equation}
\begin{gathered}
    \bs{g} = \psi_1(r) \bs{q} +\! \psi_2(r)\bs{\alpha}^2 +\!\psi_{3}(r)\bs{\mathrm{d}}r^2 +\!\psi_4(r)\bs{\alpha}\vee\bs{\mathrm{d}}r\;,
\end{gathered}
\end{equation}
with $\bs{q}$ and $\bs{\alpha}$ defined in \eqref{eq:qalph}. 
Due to the residual diffeomorphism freedom, we can further fix a gauge in which ${\psi_3=1/\psi_2}$ and  ${\psi_4=0}$. Denoting the two remaining functions ${a=\psi_2}$ and ${c=\psi_1}$, one arrives at \eqref{eq:metric_ac}.
It is also useful to perform the change of coordinates ${(\tau,\rho,\chi)\to(u,w,\phi)}$ so that $\bs{q}$ and $\bs{\alpha}$ becomes \eqref{eq:met_wu}, where the corresponding transformation reads
\begin{gather}
    u= \tfrac{\rho \sin \tau +1}{\sqrt{\rho^2-1}-\rho \cos \tau }\;, \quad w = \rho \cos \tau -\sqrt{\rho^2-1}\;,\nonumber
    \\
    \phi=\chi+2 n \log \left(\rho \cos \tau-\sqrt{\rho^2-1}\right)+4 n \tanh ^{-1}\left(\tan \tfrac{\tau }{2}\right)\nonumber
    \\
    -4 n \tanh ^{-1}\left[\csc \tau-\cot \tau \tanh \left(\tfrac{1}{2} \cosh ^{-1}\rho \right)\right]\;.
\end{gather}

The conformal form \eqref{eq:met_OmegaH} is analogous to that introduced in the static spherically symmetric case in \cite{Podolsky:2019gro}. It can be further generalized to metric ansatzes admitting the symmetries of A-metrics (A), B-metrics (B), Taub--NUT spacetimes (TNUTs), and swirling universe (SU), which take the form
\begin{equation}
\begin{aligned}
    \bs{g}_{\textrm{A/TNUTs}} &=\Omega^2(\bar{r})\left(-H(\bar{r}) \hat{\bs{\alpha}}_k^2+\tfrac{\bs{\mathrm{d}}\bar{r}^2}{H(\bar{r})}+\hat{\bs{q}}_k\right)\;,
    \\
    \bs{g}_{\textrm{B/NHEK/SU}} &=\Omega^2(\bar{r})\left(+H(\bar{r}) \check{\bs{\alpha}}_k^2+\tfrac{\bs{\mathrm{d}}\bar{r}^2}{H(\bar{r})}+\check{\bs{q}}_k\right)\;,
\end{aligned}
\end{equation}
where we denote (with the ${k=0}$ understood in the limit)
\begin{align}
    \hat{\bs{q}}_{k} &=\tfrac{\bs{\mathrm{d}}\rho^2}{1-k\rho^2}+\rho^2\bs{\mathrm{d}}\varphi^2\;,
    &
    \hat{\bs{\alpha}}_{k}&=\bs{\mathrm{d}}t+2n\tfrac{1-\sqrt{1-k\rho^2}}{k}\bs{\mathrm{d}}\varphi\;,\nonumber
    \\
    \check{\bs{q}}_{\pm1}&=\pm\rho^2\bs{\mathrm{d}}\tau^2\mp\tfrac{\bs{\mathrm{d}}\rho^2}{\rho^2-1}\;,
    &
    \check{\bs{\alpha}}_{\pm1}&=\bs{\mathrm{d}}\chi-2n\sqrt{\rho^2-1}\,\bs{\mathrm{d}}\tau\;,\nonumber
    \\
    \check{\bs{q}}_{0}&=-\bs{\mathrm{d}}\tau^2+\bs{\mathrm{d}}x^2\;,
    &
    \check{\bs{\alpha}}_{0}&=\bs{\mathrm{d}}\chi+2nx\,\bs{\mathrm{d}}\tau\;.
\end{align}
The corresponding quadratic gravity solutions with all these symmetries will be studied elsewhere \cite{GiacchiniKolarInPrep}.

\section{Recurrence relations}\label{app:recur}%
For solutions $[0,0]$ [see~\eqref{eq:OmF-series}], Eq.~\eqref{eq:EW_trFE} can be solved order by order for
\begin{align}
    f_{j+1} &=-\tfrac{\mathop{\textstyle\sum}\limits_{i=1}^{j+1}[i^2-i(6j+1)+6j(j+1)]h_i f_{j+1-i}-2\mathop{\textstyle\sum}\limits_{i=0}^{j-1}h_i f_{j-1-i}}{6j(j+1)h_0}\nonumber
    \\
    &\feq-\tfrac{ f_{j-1}}{3j(j+1)h_0}\;, \qquad j=1,2,3,\ldots\;,\label{eq:recurr00f}
\end{align}
whereas the field equation~\eqref{eq:EW_FE-tr} yields
\begin{align}
    h_4 & = \tfrac{s(6 f_1^2-6 f_0 f_2 + 3 f_0^2) - 8 (5 h_2 + 8h_0 -2) }{24 }\;, 
    \\
    h_{j+3} &=\tfrac{3s\mathop{\textstyle\sum}\limits_{i=0}^{j-1} f_i f_{j-1-i}-3s\mathop{\textstyle\sum}\limits_{i=0}^{j}(j-3i)(j+1-i)f_i f_{j+1-i}}{j(j+1)(j+2)(j+3)} \nonumber
    \\
    &\feq-\tfrac{4\left[16h_{j-1}+5j(j+1)h_{j+1}\right]}{j(j+1)(j+2)(j+3)}\;, \qquad j=2,3,\ldots\;, \label{eq:recurr00h}
\end{align}
where ${s=\sgn\omega}$. In addition, \eqref{eq:EW_FE-rr} yields a constraint between the otherwise free parameters, namely,
\begin{align}
    &3s f_0^2 \left(1- h_0 \right)+2   [\left(1-4 h_0 \right)^2-h_2^2] +3s f_1(3f_1 h_0+f_0 h_1)\nonumber
    \\
    &\qquad+2 h_1  \left(5 h_1 +3 h_3\right) = 0\;.\label{eq:const00}
\end{align}
Once this constraint is used to fix one of the coefficients, Eqs.~\eqref{eq:recurr00f} and~\eqref{eq:recurr00h} constitute the recurrence relations, which can be applied alternately to generate the analytic solution in the class $[0,0]$.
A generic solution in this class has, in principle, a total of seven free parameters: $n$, $\bar{r}_0$ and five among $\lbrace f_0, f_1, h_0, h_1, h_2, h_3\rbrace$.
The particularization for even solutions expanded around the equator $\bar{r}_0=0$ automatically fixes $f_1=h_1=h_3=0$, and~\eqref{eq:const00} can be solved for $h_0$, see~\eqref{h0pm}.

For solutions in the class $[0,1]$,  the trace~\eqref{eq:EW_trFE} of the field equations can be solved order by order for
\begin{align}
    f_1 & = -\tfrac{f_0 (1+ h_1)}{3 h_0}\;, \label{eq:recurr01f:f1}
\\
    f_{j+1} & = -\tfrac{\mathop{\textstyle\sum}\limits_{i=1}^{j+1} \left[i^2-i (6 j+5)+6 (j+1)^2\right] h_i f_{j+1-i}-2 \mathop{\textstyle\sum}\limits_{i=0}^{j-1} h_i f_{j-1-i}}{6 (j+1)^2h_0 } \nonumber
    \\
    &\feq -\tfrac{2 f_j}{6 (j+1)^2h_0 }\;, \quad j=1,2,3,\ldots\;, \label{eq:recurr01f}
\end{align}
while~\eqref{eq:EW_FE-tr} gives
\begin{align}
    h_3 & = \tfrac{s(6 f_1^2-6 f_0 f_2+3 f_0^2) -8  \left(5 h_1-2\right)  }{24 }\;,
    \\
    h_{j+2} & = \tfrac{3 s \mathop{\textstyle\sum}\limits_{i=0}^{j-1} f_i f_{j-1-i} - 3s \mathop{\textstyle\sum}\limits_{i=0}^j (j-3 i) (j+1-i) f_i f_{j+1-i} }{j (j+1) (j+2) (j+3) } \nonumber
    \\
    &\feq -\tfrac{ 4 [16 h_{j-2} + 5 j (j+1) h_j ] }{j (j+1) (j+2) (j+3)}\;, \quad j=2,3,\ldots\;. \label{eq:recurr01h}
\end{align}
Moreover,~\eqref{eq:EW_FE-rr} yields a constraint which, in view of~\eqref{eq:recurr01f:f1}, can always be used to fix
\begin{equation}\label{eq:h201app}
    h_2 = \tfrac{ s f_0^2 \left(h_1-2\right)-2  \left(5 h_0^2-h_1^2+1\right)}{6 h_0 }\;.
\end{equation}
Hence, a generic solution in the class $[0,1]$ has a total of five free parameters: $\bar{r}_0$,  $n$, $f_0$, $h_0$ and $h_1$.

\section{Horizon geometric quantities}
\label{app:geometric}%
The expressions for the geometric quantities on the horizon section $\mathcal{H}$ used in the main text follow by direct evaluation from the induced metric $\bs{\gamma}$ and the rotation 1-form $\bs{h}$~\eqref{eq:horizondata},
\begin{equation}
\begin{aligned}
    \bs{\gamma} &=a(r)\bs{\mathrm{d}}\phi^2 + \tfrac{\bs{\mathrm{d}}r^2}{a(r)}=\Omega^2(\bar{r})\left[H(\bar{r}) \bs{\mathrm{d}}\phi^2+\tfrac{\bs{\mathrm{d}}\bar{r}^2}{H(\bar{r})}\right]\;, 
    \\
    \bs{h} &=-\tfrac{2na(r)}{c(r)}\bs{\mathrm{d}}\phi-\!\tfrac{c'(r)}{c(r)}\bs{\mathrm{d}}r=-2n H(\bar{r})\bs{\mathrm{d}}\phi-\!\tfrac{2\Omega'(\bar{r})}{\Omega(\bar{r})}\bs{\mathrm{d}}\bar{r}\;.\!\!
\end{aligned}
\end{equation}

\begin{figure}[t]
    {
    \includegraphics[width=0.45\linewidth]{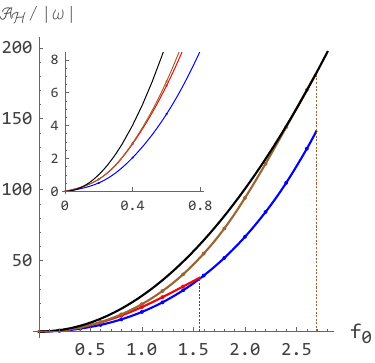}
    \hfill
    \includegraphics[width=0.45\linewidth]{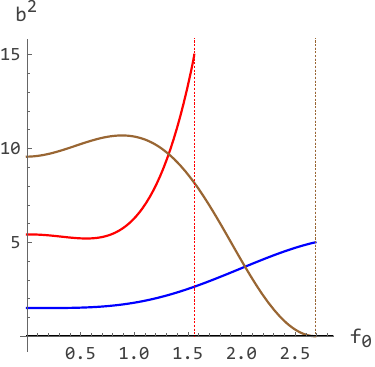}
    }
    \caption{Horizon area $\mathcal{A}_{\mathcal H}$ (left) and Bach parameter (right) for the Bachian branches of the non-tachyonic theory ($h_{0-}$ in blue, $h_{0+}$ in red), of the tachyonic theory (brown), and for NHEK (black). Dotted vertical lines indicate the termination points of the Bachian branches.}
    \label{fig:BachArea}
\end{figure}

\begin{widetext}

\begin{figure}[H]
    \includegraphics[width=0.2\linewidth]{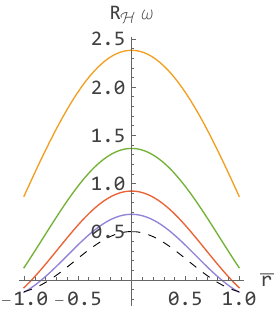}
    \hfill
    \includegraphics[width=0.2\linewidth]{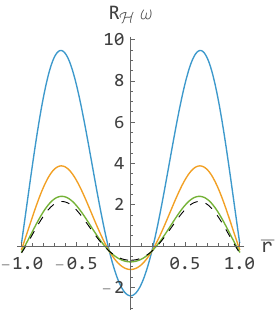}
    \hfill
    \includegraphics[width=0.2\linewidth]{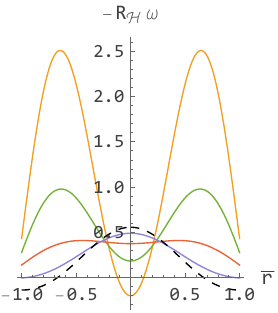}
    \hfill
    \includegraphics[width=0.2\linewidth]{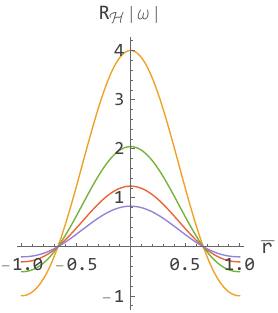}
    \\
    \subfloat[\label{subfig:a}]{
    \includegraphics[width=0.2\linewidth]{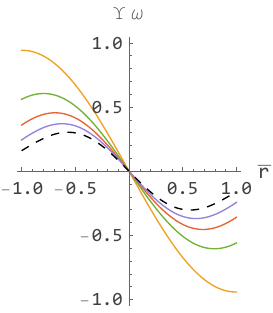}
    }
    \hfill
    \subfloat[\label{subfig:b}]{
    \includegraphics[width=0.2\linewidth]{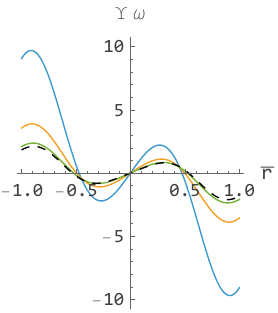}
    }
    \hfill
    \subfloat[\label{subfig:c}]{
    \includegraphics[width=0.2\linewidth]{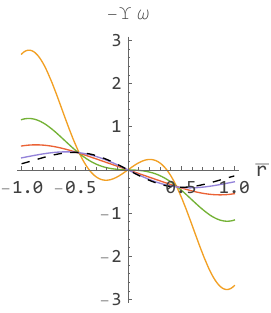}
    }
    \hfill
    \subfloat[\label{subfig:d}]{
    \includegraphics[width=0.2\linewidth]{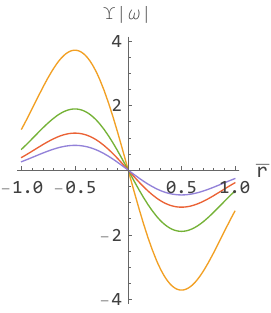}
    }
    \\
    \includegraphics[width=0.4\linewidth]{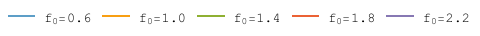}
    \caption{${R_{\mathcal{H}}}$ and $\Upsilon$ for the Bachian branches $h_{0-}$ (a) and $h_{0+}$ (b) of the non-tachyonic theory, for the branch of the tachyonic theory (c), and for the NHEK branch (d), for several values of ${f_0=\Omega(0)/\sqrt{|\omega|}}$. The dashed curves indicate the limiting solution for ${f_0\approx 2.69}$ (a,c) or ${f_0\approx 1.56}$ (b).}
    \label{fig:AP}
\end{figure}
\end{widetext}

The horizon area $\mathcal{A}_{\mathcal H}$ (which measures the overall size of the section), the scalar curvature $R_{\mathcal{H}}$ of the induced metric (characterizing its deformation), and the rotational scalar $\Upsilon$ (encoding the rotational structure of the horizon~\cite{Ashtekar:2004cn}) are given by
\allowdisplaybreaks
\begin{align}
    \mathcal{A}_{\mathcal H} &=\mathop{\textstyle\int}\limits_{\mathcal H}\sqrt{\det\bs{\gamma}} =-\tfrac{8\pi r_*}{a'(r_*)}=-\tfrac{8\pi}{H'(\bar r_*)}\mathop{\textstyle\int}\limits_0^{\bar r_*}\Omega^2(\bar r)\,\mathrm{d}\bar r\;,\nonumber
    \\
    R_{\mathcal{H}}&=-a''=
\tfrac{1}{\Omega^2}\big[-H''-2\big((\log\Omega)'H\big)'\big]\;,
    \\
    \Upsilon&=*\bs{\mathrm{d}}\bs{h}=2n\left(\tfrac{a}{c}\right)'=2n\tfrac{H'}{\Omega^2}\;,\nonumber
\end{align}
where $*$ and $\bs{\mathrm{d}}$ denote the Hodge dual and exterior derivative on $\mathcal{H}$, respectively. (Note that $\Upsilon$ is related to the Newman--Penrose scalar $\Psi_2$ in the frame of \cite{Colleaux:2026} by ${\Upsilon=-4\Im\Psi_2}$.) For the NHEK branch, these quantities take the form
\begin{equation}
\begin{aligned}
    \mathcal{A}_{\mathcal H}&=8\pi\Omega_0^2\;,\\
    R_{\mathcal{H}}&=-\tfrac{4}{\Omega_0^2}\cos^4(n\bar r)\big(2\cos(2n\bar r)-1\big)\;,\\
    \Upsilon&=-\tfrac{4}{\Omega_0^2}\cos^3(n\bar r)\sin(3n\bar r)\;.
\end{aligned}
\end{equation}
The above horizon quantities are plotted for the NHEK and Bachian branches in Figs.~\ref{fig:BachArea} and~\ref{fig:AP}.

Also, in Fig.~\ref{fig:BachArea} we plot the Bach parameter $b^2$, defined as proportional to the value of $B_{ab}B^{ab} = [\mathcal{B}_1^2 + 2 (\mathcal{B}_1 + \mathcal{B}_2)^2]/(72 \Omega^8)$ [see~\eqref{eq:Bach12}] at the equator, namely
\begin{equation}
   b^2 = \mathcal{B}_1^2 + 2 (\mathcal{B}_1 + \mathcal{B}_2)^2 \vert_{\bar{r}=0} \; .
\end{equation}
As expected, for the Bachian branch of the tachyonic theory, $b^2 \to 0$ as $f_0$ approaches its upper limit, where this branch meets the NHEK one.

\vfill

\end{document}